# Assessment of sol-gel derived iron oxide substituted 45S5 bioglass-ceramics for biomedical applications.

Nitu [a], Rushikesh R Fopase [b], Lalit Mohan Pandey [b], Papori Seal [c], Jyoti Prasad Borah [c] and Ananthakrishnan Srinivasan*[a]

*Department of Physics, Indian Institute of Technology Guwahati, Guwahati-781039, India.*

*Department of Biosciences and Bioengineering, Indian Institute of Technology Guwahati, Guwahati-781039, India.*

*Department of Science and Humanities, National Institute of Technology Nagaland, India 797103, India.*

**Abstract**

Magnetic bioactive glass-ceramic (MGC) powders with nominal compositions of $(45-x)SiO_2$ $24.5CaO$ $24.5Na_2O$ $6P_2O_5$ $xFe_2O_3$ (x = 2, 4, 6, 8, 10, 15 wt. %) have been synthesized by sol-gel route by systematically substituting silicon dioxide with iron oxide in the Hench's 45S5 glass composition. Powder x-ray diffraction studies revealed a variation in the percentage of combeite ($Ca_2Na_2Si_3O_9$), magnetite ($Fe_3O_4$), and hematite ($Fe_2O_3$) nanocrystalline phases in MGC powders as a function of composition. Zeta potential measurements showed that MGC containing up to 10 wt.% iron oxide formed stable suspensions. Saturation magnetization and heat generation capacity of MGC fluids increased with increase in iron oxide content. Degradation of MGC powders was investigated in phosphate buffer saline (PBS). In vitro bioactivity of the MGC powders taken in pellet form was confirmed by observing the pH variation as well as hydroxyapatite layer (HAp) formation upon soaking in modified simulated body fluid. These studies showed a decrement in overall bioactivity in samples with high iron oxide content due to the proportional decrease in silanol group. Monitoring the proliferation of MG-63 osteoblast cell in Dulbecco's Modified Eagle Medium (DMEM) revealed that MGC with up to 10 wt.% iron oxide exhibited acceptable viability. The systematic study revealed that the MGC with 10 wt.% iron oxide exhibited optimal cell viability, magnetic properties and induction heating capacity which were better than those of FluidMag-CT, which is used for hyperthermia treatment.

## 1. Introduction

Glass with composition $45SiO_2 24.5CaO 24.5Na_2O 6P_2O_5$ (wt.%) popularly known as the 45S5 Bioglass® or the Hench's 45S5 bioglass (in honor of its inventor L L Hench) is a synthetic biomaterial which makes a bond with the human bone and stimulates bone growth in the presence of the body fluid.[1] This invention ushered in a broad range of new bioactive materials and ceramics such as synthetic hydroxycarbonate apatite, calcium sulfate and apatite wollastonite glass-ceramics.[2,3] A small bone replacement involving middle ear surgery is the first successful bioglass implant in a human being.[4] A variety of bioactive glasses and glass-ceramics have been synthesized with different constitutions and compositions in the search for an ideal implant material.[5-7] Despite these efforts spanning over the last four decades, the 45S5 bioglass remains the best bioactive glass composition.[7] Melt quenching and sol-gel techniques are the most preferred methods for the synthesis of bioactive glasses and glass-ceramics in bulk and powder forms, respectively. The sol-gel derived bioactive glass powder has several advantages over its bulk counterpart such as being much cheaper, easier to process in large quantities, lower in impurity and better in compositional homogeneity.[8] Sol-gel processing of three component $CaO-SiO_2-P_2O_5$ glass is relatively easy. However, it is not a fully successful technique for preparing glasses with high $Na_2O$ content due to the high hydrolytic reactivity of sodium alkoxide in water.[9] Considering the enhanced bioactivity of sol-gel derived glasses as compared to their bulk counterparts, a few attempts have been made to synthesize $CaO-SiO_2-P_2O_5-Na_2O$ glasses using organic solvents such as ethylene glycol, formic acid, and acetic acid.[10-12] But, these synthesis protocols are too tedious and time consuming for easy adaption and commercialization. Moreover, attempts to prepare sol-gel based quaternary glasses with high $Na_2O$ content invariably result in the crystallization of combeite ($Na_2Ca_2Si_3O_9$) phase, leading to the devitrification of the glassy matrix into a glass-ceramic.[9,13] It is worthy to point out that bulk 45S5 glass-ceramic with $Na_2Ca_2Si_3O_9$ phase has better mechanical strength than bulk 45S5 Bioglass without degradation in bioactivity and hence serves as a better implant material.[14,15]

Hyperthermia treatment (HT) of cancer cells using a magnetic bioactive implant as a thermoseed is considered to be a compelling post-tumor removal procedure, especially, in deep rooted bone tumors.[16] In such a treatment, the crevice left after the removal of the malignant tumor is filled with the magnetic bioactive implant for strengthening the bone as well as for regeneration of the bone. Application of an alternating magnetic field externally, induces local heating in the vicinity of the implant site due to magnetic hysteresis loss in the magnetic implant. Periodic and controlled local heating of the thermoseed can selectively destroy the thermally susceptible malignant cells in its vicinity without destroying the thermally stable healthy cells. Though 45S5 bioglass and glass-ceramic have excellent bioactivity and biocompatibility, they are not magnetic and hence do not exhibit magnetic hysteresis under an alternating magnetic field. One can address this handicap by introducing oxides of ferromagnetic transition metals Fe, Co, and Ni in the Hench's glass composition. Of the three oxides, iron oxide is biocompatible and less cytotoxic to the human body than Co or Ni oxide. The pioneering work on using magnetic material for HT is attributed to Gilchrist et al. who also explored different procedures, magnetic materials, frequencies, and encapsulation methods.[17-19] A survey of the literature reveals that magnetite ($Fe_3O_4$) and maghemite ($\gamma$-$Fe_2O_3$) are the two forms of iron oxide which have been widely investigated for biomedical applications.[19-23] The first bulk magnetic glass-ceramic of composition $Fe_2O_3-P_2O_5-Li_2O-SiO_2-Al_2O_3-MnO-B_2O_3-MgO$ for HT was patented by Borrelli et al. in 1981.[24] Thereafter, Luderer et al. proposed the use of bulk $11.6Li_2O-0.4Al_2O_3-3.4SiO_2-23.7P_2O_5-60.5Fe_2O_3$ glass-ceramic with a saturation magnetization ($M_s$) of 8.69 emu/g for HT of cancer.[25] This specimen which yielded a specific absorption rate (SAR) of ~1 W/g under an applied field of 500Oe and frequency of 10 kHz could raise the local temperature rise to 43.5 °C in seven minutes when tested against breast carcinoma tumor in a subcutaneous rat model and led to the full cure of ~12% of the animals. Kukubo et al. developed a $P_2O_5$-free bulk glass-ceramic of composition $40Fe_2O_3-60(CaO-SiO_2)$ (wt.%) and demonstrated its bioactivity and magnetic properties.[26] Ebisawa et al. reported temperature dependent magnetic properties of this sample and the influence of small amounts of $Na_2O$, $B_2O_3$, and $P_2O_5$ addition in its bioactivity.[27-29] Levenouri et al. investigated the structural and magnetic properties of bulk $CaO-SiO_2-P_2O_5-Na_2O-Fe_2O_3$ glass-ceramic.[30] They also studied influence of preparative conditions on the magnetic properties and heat generation of $CaO-SiO_2-P_2O_5-Na_2O-Fe_2O_3$ glass-ceramic prepared by melt quenching and co-precipitation methods.[31] Later, Bretcanu et al. interpreted the magnetic and induction-heating behavior of co-precipitated bulk $24.7SiO_2-3.3P_2O_5-13.5CaO-13.5Na_2O-14FeO-31Fe_2O_3$ glass-ceramic.[32] Singh et al. performed a systematic study of bulk $(45-x)CaO-34SiO_2-16P_2O_5-4.5MgO-0.5CaF_2-xFe_2O_3$ ($x$ = 5–20 wt.%) ferrimagnetic bioglass-ceramics and found hydroxyapatite, magnetite, and wollastonite as major crystalline phases in all compositions.[33] Their study on bulk 41CaO-(52-x)SiO$_2$-4P$_2$O$_5$-xFe$_2$O$_3$-3Na$_2$O (x = 0, 2, 4, 6, 8 and 10 mol%) glass-ceramics showed that $M_s$ ($H$c) varied from 0.17 emu/g (523 Oe) to 7.95 emu/g (91 Oe) as x was varied from 0 to 10 mol%.[34]

It is evident from the literature that iron oxide substituted Hench's 45S5 bioglass-ceramic has not been well explored for use as in HT and sol-gel derived form of this glass-ceramic has distinct advantages. This motivated us to synthesize magnetic bioglass-ceramics with bulk 45S5 Bioglass composition by sol-gel route and evaluate the influence of progressive iron oxide substitution for silicon oxide on its bone regeneration and HT related properties. We present a systematic study of structural, magnetic, heat generation and *in vitro* bioactivity of the iron oxide substituted Hench's 45S5 glass-ceramics (MGC) as a function of iron oxide content and an assessment of their application in bone regeneration and HT of cancer.

## 2. Materials and methods

### 2.1. Synthesis

Magnetic bioactive glass-ceramic powders (MGC) of compositions (45-x)SiO$_2$ 24.5CaO 24.5Na$_2$O 6P$_2$O$_5$ xFe$_2$O$_3$ (x = 2, 4, 6, 8, 10, 15 wt. %) have been synthesized through the sol-gel route. The compositions and nomenclature of the synthesized samples are listed in Table 1. Tetraethyl orthosilicate [TEOS, Si(OC$_2$H$_5$)$_4$, purity > 99%, Sigma-Aldrich], calcium nitrate tetrahydrate [Ca(NO$_3$)$_3$.4H$_2$O, purity > 98%, Loba Chemie], triethyl phosphate [TEP (C$_6$H$_{15}$O$_4$P), purity > 99%, Sigma-Aldrich), sodium nitrate (NaNO$_3$, purity > 98%, Loba Chemie) and iron nitrate [Fe(NO$_3$)$_3$. 9H$_2$O, purity > 98%, Loba Chemie] were used as sources of SiO$_2$, CaO, P$_2$O$_5$, Na$_2$O, and iron oxide, respectively. TEOS:H$_2$O molar ratio of 1:18 was maintained while preparing the samples. A solution of HNO$_3$ (1M) and distilled water was first prepared. Then, TEOS was slowly added to aqueous HNO$_3$ under constant stirring. After allowing for completion of hydrolysis of TEOS, other precursors were introduced sequentially into the TEOS sol. The sol was then left for aging at room temperature for 3 days. The dried gel powder was kept in an air oven at 70 °C for 3 days for thermal stabilization. Finally, the dried powder was heat treated at 750 °C for 2 hours and the resultant product used for further studies.

Table 1 Nominal composition calculated crystalline phase percentages in various samples.

| Sample Code | Composition (wt.%) | | | | | Crystalline phases present (%) | | |
|---|---|---|---|---|---|---|---|---|
| | SiO$_2$ | CaO | Na$_2$O | P$_2$O$_5$ | Fe$_2$O$_3$ | Combeite | Hematite | Magnetite |
| S$_{0Fe}$ | 45 | 24.5 | 24.5 | 6.0 | 0 | 100.0 | 0.0 | 0.0 |
| S$_{2Fe}$ | 43 | 24.5 | 24.5 | 6.0 | 2 | 84.3 | 6.1 | 9.6 |
| S$_{4Fe}$ | 41 | 24.5 | 24.5 | 6.0 | 4 | 80.3 | 7.1 | 12.6 |
| S$_{6Fe}$ | 39 | 24.5 | 24.5 | 6.0 | 6 | 77.0 | 8.6 | 14.4 |
| S$_{8Fe}$ | 37 | 24.5 | 24.5 | 6.0 | 8 | 70.6 | 21.6 | 27.8 |
| S$_{10Fe}$ | 35 | 24.5 | 24.5 | 6.0 | 10 | 46.3 | 24.8 | 29.7 |
| S$_{15Fe}$ | 30 | 24.5 | 24.5 | 6.0 | 15 | 22.7 | 34.6 | 42.7 |

## 2.2. Characterization

The Brunauer-Emmett-Teller (BET) method was used to calculate surface area of the sol-gel derived powders from physisorption isotherms recorded at 77 K using Quntachrome Autosorb-IQ MP instrument. A set of 10 experimental points were chosen in a linear region of the data $0.05 < P/P_0 < 0.3$ (where $P$ is the applied gas pressure and $P_0$ is the relative gas pressure) to determine the surface area. Powder XRD patterns of the powders were recorded using the Cu K$_\alpha$ X-ray line ($\lambda = 1.5406$ Å) of a Rigaku TRAX III powder X-ray diffractometer operating at 5 kW. The surface morphology and elemental composition of the powders were analysed using a scanning electron microscope (SEM, ΣIGMA, ZEISS) equipped with an energy dispersive spectrometer (EDS) unit. Field emission transmission electron microscope (FETEM, JEOL-2100F) was employed to record selected area electron diffraction (SAED) and high resolution TEM (HRTEM) images of the powder particles. The zeta potential ($\zeta$), an electrokinetic parameter which is determined by the surface charge of colloidal particle in suspension, was calculated using the expression (1). [35]

$$Zeta\ potential\ (\zeta) = \frac{3\eta\mu_e}{2\varepsilon f(k\ a)} \qquad (1)$$

In this expression, $\eta$ is the viscosity of the suspension, $\mu_e$ is the mobility of colloidal particle, $\varepsilon$ is the dielectric constant of the solution, $f(k\ a)$ is the Henry function, $k$ is Debye length (whose inverse gives the thickness of the charge layer), and $a$ is the particle equivalent spherical diameter. The $\zeta$ of sonicated MGC powder was measured at 25 °C using a Zetasizer (ZEN3600, Malvern). 1 mg of each MGC powder was dispersed in 1 ml of deionized water with pH of 6.8 and sonicated for 30 minutes before the $\zeta$ measurement. Room temperature magnetic measurement was performed with a physical property measurement system (PPMS, Quantum Design, DynaCool) based vibrating sample magnetometer (VSM). Easy Heat 8310 (Ambrell, UK) was used to evaluate the heat generation capacity of MGC powders dispersed in deionized water. Temperature changes in the magnetic fluids were recorded up to 900 seconds using a near infrared region imaging camera. The degradation of the powder samples was monitored in phosphate buffer saline (PBS) of pH 7.4 at 37 °C. Degradation of the samples was evaluated by measuring the weight loss of the MGC powder pellet before and after immersion in PBS of pH 7.4 at 37 °C for different soaking time (1, 3 and 5 days) using the relation (2),

$$Weight\ loss\ (\%) = \frac{w_0 - w_t}{w_0} \times 100\ \% \qquad (2)$$

In relation (2), $w_0$ and $w_t$ are the weights of the sample before and after immersion for a particular time in PBS, respectively. *In vitro* bioactivity test was carried out using pelletized powders soaked in modified simulated body fluid (SBF) of pH of 7.40 maintained at 37 °C by monitoring the pH and HAp layer formation as a function of time.[36] Osteoblast MG-63 cells were cultured using DMEM (pH 7.4) in $CO_2$ incubator at 37 °C and 85% humidity. To perform this test, 5000 cells were added to each well of a 96-well culture plate. The plates were incubated in a $CO_2$ incubator for 01, 03 and 05 days. After the completion of the incubation periods, the media from the wells was replaced with fresh media along with 2,5-diphenyl-2H-tetrazolium bromide (MTT) (5 mg/ml) followed by 4 hours of incubation. During the incubation, the added MTT reduces to formazan crystals through oxidoreductase enzymes secreted by live cells in the wells. The formed formazan crystals then dissolved in the dimethyl sulfoxide (DMSO) and its absorbance was recorded at the wavelength of 570 nm using a multi-plate reader (Fluostar Omega, BMG, Germany). The cell viability was estimated using the relation (3),

$$Cell\ viability\ (\%) = \frac{Absorbance\ of\ Sample\ at\ 570\ nm}{Absorbance\ of\ Control\ at\ 570\ nm} \times 100 \quad (3)$$

## 3. Result and discussion

### 3.1. Surface area analysis

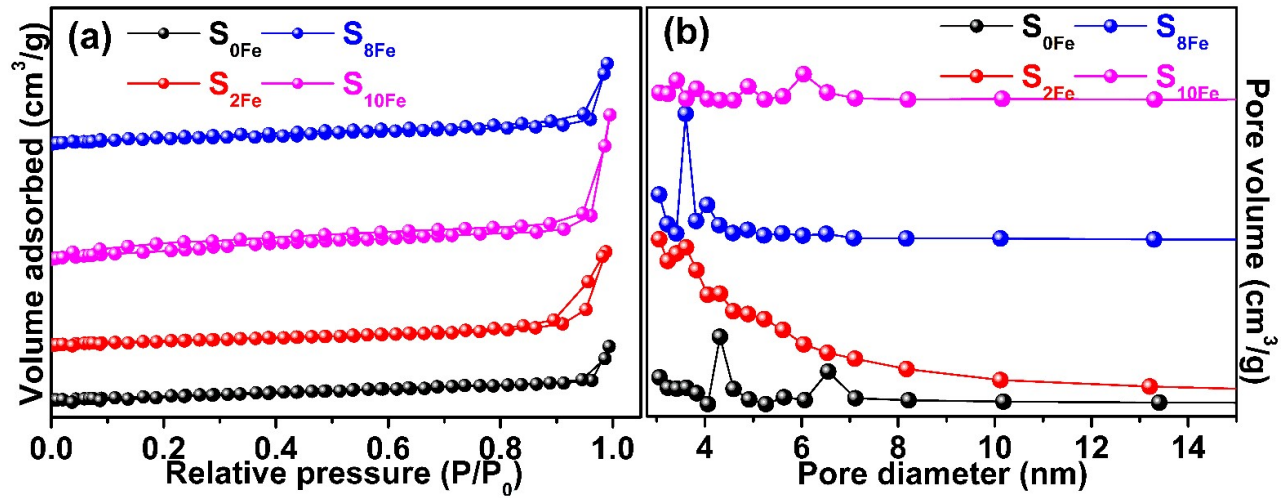

Fig. 1. Nitrogen adsorption-desorption (a) isotherms and (b) pore size distribution curves of samples $S_{0Fe}$, $S_{2Fe}$, $S_{8F}$, and $S_{10Fe}$.

The textural properties of the samples were inferred using $N_2$ adsorption-desorption isotherms which were analyzed using theoretical models prescribed by the International Union of Pure and Applied Chemistry. Fig. 1 (a) and (b) show the isotherms and pore size distributions of the sol-gel derived MGC powders. All the powder samples exhibited H2-type isotherm which is a part of isotherm type-IV representing a mesoporous interconnected worm-like structure. It is also evident from the data presented that progressive substitution of iron oxide for silica in the nominal 45S5 composition does not induce any significant change in the shape of the isotherm. The specific surface area, mean pore size and total pore volume calculated using the BET and Barrett-Joyner-Halenda (BJH) methods are presented in Table 2. A significant variation in the surface area, average pore size, and pore volume are observed in the samples with an increase in iron oxide. The pore size distribution was obtained from the desorption isotherm by the BJH method. Mono-modal type of distribution is observed for the $S_{2Fe}$ and $S_{8Fe}$ samples, while bi-modal type distribution has been observed for $S_{0Fe}$ and $S_{10Fe}$ (Fig.1 (b)). The latter could have some advantage in osteointegration process involving proteins of different sizes.[37] The average pore size and total pore volume increased from 3.31 nm to 4.18 nm, and 0.012 cc/g to 0.031 cc/g, respectively, when iron oxide was increased from 0 to 8 wt.%. Though $S_{10Fe}$ has the highest surface area among the samples, its average pore size and total pore volume are close to that of $S_{0Fe}$ sample. As the iron oxide content was increased from 0 to 10 wt.%, surface area of the glass-ceramic powder increased from 1.81 m$^2$/g to 4.03 m$^2$/g. S$_{0Fe}$, which is devoid of iron oxide shows the minimum surface area of 1.81 m$^2$/g.

Table 2 Textural properties of S$_{0Fe}$, S$_{2Fe}$, S$_{8Fe}$, and S$_{10Fe}$.

| Sample code | Surface area (m$^2$/g) | Pore size (nm) BJH | Pore volume (cc/g) |
|---|---|---|---|
| S$_{0Fe}$ | 1.81 | 3.31 | 0.012 |
| S$_{2Fe}$ | 3.81 | 4.18 | 0.020 |
| S$_{8Fe}$ | 3.86 | 4.18 | 0.031 |
| S$_{10Fe}$ | 4.03 | 3.72 | 0.017 |

The measured value of surface area shows slightly higher value with the reported value of 0.9 m$^2$/g for nanopowders of 45S5 glass-ceramic in the literature.[38] As iron oxide was gradually added to the Hench's 45S5 composition in the place of silicon oxide, the specific area of the glass-ceramic increased. This increase in the surface area is due to the larger size of iron ions as compared to Si ions. It can be observed from the experimental data that the porosity of the glass-ceramic is highly influenced by the silica content of the material. On the other hand, the iron oxide content enhances the surface area and influences the porosity of the sol-gel derived powder. A survey of the literature shows that this is the study on the influence of iron oxide on the surface properties of sol-gel derived powders prepared without template. The overall enhancement of textural properties of the mesoporous S$_{10Fe}$ shows great promise for biomedical applications of this iron oxide substituted 45S5 glass-ceramic.

### 3.2. Structural analysis

X-ray diffraction (XRD) patterns of MGC powders designated as S$_{0Fe}$, S$_{2Fe}$, S$_{4Fe}$, S$_{6Fe}$, S$_{8Fe}$, S$_{10Fe}$, and S$_{15Fe}$ are shown in Fig. 2 (a). Three crystalline phases, *viz.*, combeite (Na$_2$Ca$_2$Si$_3$O$_9$, ICDD 002-1445), hematite ($\alpha$-Fe$_2$O$_3$, ICDD 1-072-0469), and magnetite (Fe$_3$O$_4$, ICDD00-019-0629) were identified in all the powders with iron oxide. The percentage of each crystalline phase estimated from the XRD data is mentioned in Table 1. It can be observed from Table 1 that the combeite phase is more pronounced in those samples having a low concentration of iron oxide. Iron oxide phases (magnetite and hematite) were found to significantly increase as more iron oxide precursor was added with a corresponding decrease in combeite phase. The combeite phase has been well studied and is found to have good mechanical strength and biodegradability.[15,39,40] The magnetic properties are attributed to available magnetic phases (magnetite and hematite) in the sol-gel powders. Though the crystallization of bioactive glass tends to decrease the bioactivity, crystallization of combeite phase compensates it and balanced the decrement in bioactivity.[41] Therefore, the combined presence of combeite and the iron oxide phases ensures both biocompatibility and induction heating capacity of the iron oxide substituted MGC powders.

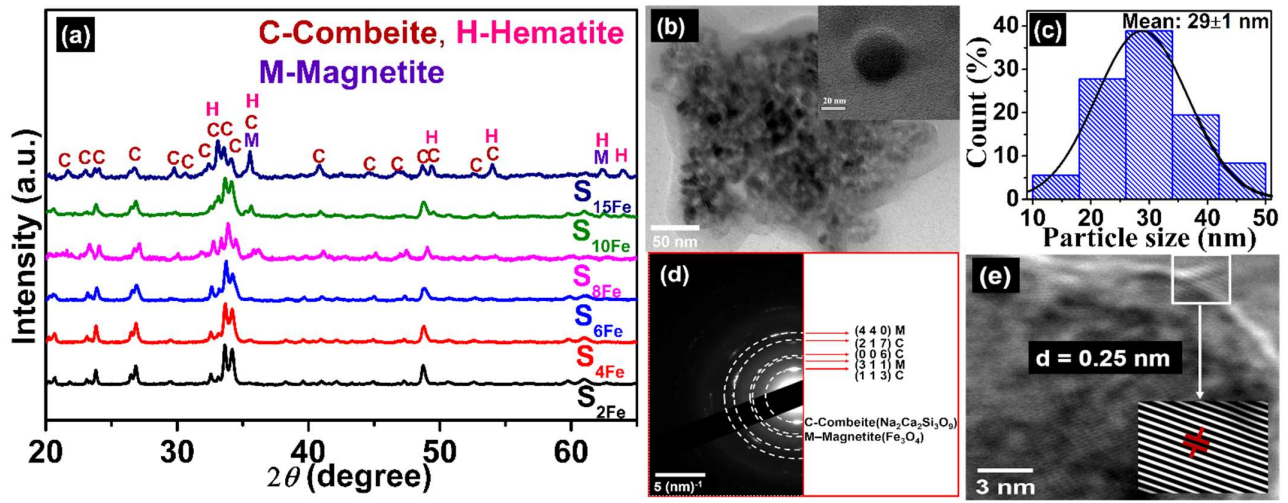

Fig. 2. (a) XRD patterns of samples $S_{0Fe}$, $S_{2Fe}$, $S_{4Fe}$, $S_{6Fe}$, $S_{8Fe}$, $S_{10Fe}$, and $S_{15Fe}$. (b) FETEM image (c) particle size distribution calculated from the TEM image (d) SAED pattern and (e) HRTEM image of $S_{10Fe}$. Inset of (e) shows iFFT of (331) plane of magnetite phase in $S_{10Fe}$.

One would expect better induction heating capacity in MGC powder with higher iron oxide content since they contain higher magnetic content. However, the decrease in combeite phase percentage with an increase in the percentages of the magnetic phases should not lead to significant decrease in the bioactivity of the glass-ceramics. Hence, it is necessary to study the magnetic, heat generation and bioactivity of the samples $S_{0Fe}$, $S_{2Fe}$, $S_{4Fe}$, $S_{6Fe}$, $S_{8Fe}$, $S_{10Fe}$, and $S_{15Fe}$ to assess their individual potential for biomedical applications.

Morphology of dispersed $S_{10Fe}$ powder particles can be visualized in the FETEM image shown in Fig. 2 (b) and the near-spherical shape of a typical particle is illustrated as an inset in Fig. 2 (b). The particle size distribution of $S_{10Fe}$ powder is shown in Fig. 2 (c) from which the average particle size was estimated to be 29 ±1 nm with a full width and half maximum (FWHM) of 16 nm. The SAED pattern of the $S_{10Fe}$ sample shows reflections from (113), (006) and (217) planes of the combeite phase and (311) and (440) reflections from magnetite phase. HRTEM image and inverse fast Fourier transform (iFFT) of HRTEM image reveal the lattice fringe corresponding (311) plane of magnetite phase as displayed in Fig. 2 (d-e).

### 3.3. Surface Zeta potential analysis

The contact between the solid surface of powder and the aqueous medium constructs the overall charge over the surface of particles.[42] This surface charge is one of the main factors which influences the hydrodynamic interaction of the particles in the medium, it plays a vital role in the interaction of the material with the biological environment in such processes as protein adsorption, and cellular and bacterial adhesion.[42] The zeta potential ($\zeta$) of aqueous suspensions of MGC powders, which is a measure of the surface charges in the fluid, was recorded and the corresponding data are displayed in Fig. 3. $\zeta$ of MGC powders decreased in magnitude from -34.7 to -7.0 as iron oxide content was increased from 2 wt.% to 15 wt.% in Hench's 45S5 composition. A general hydrokinetic stability standard classifies suspension in the $\zeta$ range of +25 mV to -25 mV as unstable and susceptible to form aggregation and agglomeration in water medium.[43] According to this classification, the $\zeta$ value of aqueous suspensions of $S_{2Fe}$, $S_{4Fe}$, $S_{6Fe}$, $S_{8Fe}$ and $S_{10Fe}$ are outside this range and hence can be classified as stable suspensions.

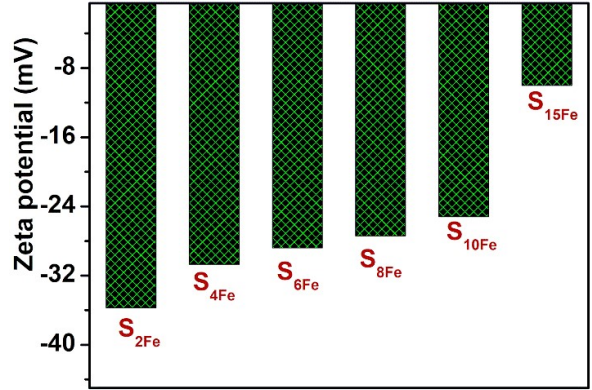

Fig. 3. Zeta potential ($\zeta$) of $S_{2Fe}$, $S_{4Fe}$, $S_{6Fe}$, $S_{8Fe}$, $S_{10Fe}$, and $S_{15Fe}$ recorded at room temperature.

However, the aqueous suspension of $S_{15Fe}$ is within the range of unstable suspensions and hence prone to aggregation and agglomeration in water medium. This reduction in $\zeta$ as a function of iron oxide can be understood on the basis of the interaction of the magnetic particles dispersed in water. Structural studies show a progressive increase in the percentage of magnetic phases with an increase in iron oxide content. Hence, with an increase in magnetic phase content, the magnetization increases, and hence the magnetic attraction between the particles. This in turn increases the aggregation and agglomeration of the particles. This leads to the destabilization of the colloidal suspension which reflects as a decrease in $\zeta$ with an increase in iron oxide content. This destabilization mechanism becomes very strong for suspensions with magnetic content higher than $S_{10Fe}$ making them unstable as observed in the case of $S_{15Fe}$ suspension.

### 3.4. Magnetic properties evaluation

Fig. 4 (a) shows the recorded room temperature magnetic hysteresis loop of $S_{0Fe}$, $S_{2Fe}$, $S_{4Fe}$, $S_{6Fe}$, $S_{8Fe}$, $S_{10Fe}$, and $S_{15Fe}$ recorded for an applied magnetic field range of ±40 kOe. Fig. 4 (b) depicts the minor hysteresis loops recorded in the clinically viable magnetic field range of ±500 Oe. Low iron oxide containing samples such as $S_{2Fe}$ and $S_{4Fe}$ did not exhibit magnetic saturation up to 40 kOe because of the relatively lower percentage of the magnetic phases present as compared to the combeite phase. As iron oxide was increased in the nominal composition, the percentage of magnetic phases increased and hence the $M_s$ of the samples. Low coercivity ($H_c$) and low remnant moment ($M_r$) values indicate the soft magnetic behavior of the MGC powders. Magnetic parameters such as $M_s$, $H_c$, $M_r$, and hysteresis loop area of each sample are listed in Table 3. $M_s$ increased from 0.095 emu/g to 1.543 emu/g on increasing iron oxide content from 2 to 15 wt.% in the Hench's 45S5 composition. Magnetic properties along with bioactivity are crucial for hyperthermia and related biomedical applications. Many studies have been carried out to explore these properties of iron oxide substituted glasses and glass-ceramics. Luderer *et al.* have reported $M_s$ of 8.69 emu/g in bulk $11.6Li_2O-0.4Al_2O_3-3.4SiO_2-23.7P_2O_5-60.5Fe_2O_3$ glass-ceramic, Kokubo *et al*. found $M_s$ of 30 emu/g in bulk $40Fe_2O_3-60(SiO_2-CaO)$ glass-ceramic, Bretcanu *et al*. obtained an $M_s$ of 34 Am$^2$/kg in bulk glass-ceramic $13.5CaO-24.7SiO_2-3.3P_2O_5-13.5Na_2O-14FeO-31Fe_2O_3$ (wt.%).[25,26,32]

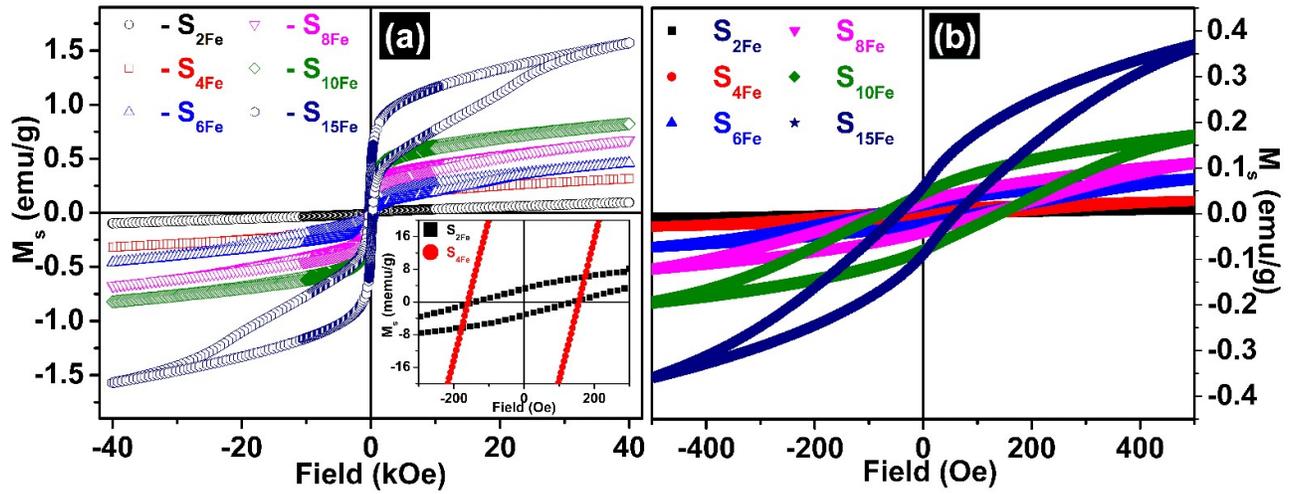

Fig. 4 Room temperature *M-H* loop of $S_{2Fe}$, $S_{4Fe}$, $S_{6Fe}$, $S_{8Fe}$, $S_{10Fe}$, and $S_{15Fe}$ recorded at (a) ±40 kOe and (b) ±500 Oe. Inset in (a) provides an enlarged view of data close to the origin.

Recently, Yazdanpanah *et al.* studied the magnetic properties of sol-gel derived bioactive glass of composition $Fe_2O_3$-$P_2O_5$-CaO-$SiO_2$ in the presence of $Li_2O$ and BaO.[44,45] They found that $Li_2O$ enhanced $M_s$ while BaO reduced it. Further, Leventouri *et al.* prepared bulk magnetic glass-ceramics with same oxide components as in the present study but with different compositions (non Hench's 45S5 glass-ceramic) and obtained $M_s$ values ranging from 1 emu/g to 18 emu/g by varying the iron oxide from 5 to 20 wt.%.[30] To the best of our knowledge, there are no reports on magnetic properties of sol-gel derived iron oxide substituted Hench's 45S5 glass-ceramics. Two magnetic phases, *viz.*, hematite and magnetite, have already been identified in these MGC samples (*cf.* Fig. 2 and Table 1). The hematite phase (α-$Fe_2O_3$) exhibits antiferromagnetic nature below 260 K and shows a weak ferromagnetic behavior at room temperature (bulk $M_s$ ~ 0.4 $Am^2Kg^{-1}$), while the magnetite ($Fe_3O_4$) phase shows soft ferromagnetic nature (bulk $M_s$ ~ 92 $Am^2Kg^{-1}$).[46]

Table 3 Magnetic parameters of MGC.

| Magnetic and structural parameters | $S_{2Fe}$ | $S_{4Fe}$ | $S_{6Fe}$ | $S_{8Fe}$ | $S_{10Fe}$ | $S_{15Fe}$ |
|---|---|---|---|---|---|---|
| Saturation magnetization, $M_s$ (emu/g) | 0.094 | 0.246 | 0.447 | 0.635 | 0.820 | 1.543 |
| Coercive field, $H_c$ (Oe) | 137 | 234 | 306 | 214 | 149 | 107 |
| Remnant magnetization, $M_r$ (emu/g) | 0.0032 | 0.0213 | 0.0685 | 0.1124 | 0.1169 | 0.1247 |
| Hysteresis loop area at ±40 kOe (erg/g) | 57 | 725 | 2280 | 4352 | 8966 | 18041 |
| Hysteresis loop area at ±500 Oe (erg/g) | 2 | 8 | 22 | 35 | 62 | 74 |

Hence, the overall $M_s$ of the MGC will depend the relative percentages of the two magnetic phases. A compositional study of magnetic parameters shows that $H_c$ increases with increase in iron oxide concentration up to 6 wt.% iron oxide ($S_{6Fe}$) and then decreases for higher iron oxide concentration, whereas $M_r$ increases with iron oxide concentration. A significant variation of the *M-H* loop area is observed with increase in iron oxide content. The hysteresis loop area is directly related to the energy loss responsible for the heat generated by the MGC under an alternating magnetic field. The magnetic parameters of all the MGC listed in Table 3 indicate that the samples with high iron oxide content, *viz.*, $S_{10Fe}$ and $S_{15Fe}$ are better suited for hyperthermia applications. Since a high magnetic field is not viable for clinical *in vivo* induction heating, the *M-H* loops have also been recorded at an amenable field range of ±500 Oe. A similar trend of magnetic parameters such as $M_s$, $H_c$, $M_r$, and hysteresis loop area has been observed for the *M-H* loop recorded at 40kOe as well as 500 Oe.

### 3.5. Induction Heating

Induction heating test was performed to assess the applicability of the MGC samples as thermoseed for hyperthermia application. Considering their high $M_s$ and $H_c$, $S_{10Fe}$ and $S_{15Fe}$ were chosen for the test. The rise in temperature as a function of time and NIR camera images corresponding to three different concentrations of aqueous $S_{10Fe}$ and $S_{15Fe}$ magnetic fluid are shown in Fig. 5. Care was taken to ensure that the safety limit, expressed as a product of the applied field strength and the frequency of the alternating magnetic field (*H\*f*) value of 9.59 × $10^9$ Am$^{-1}$s$^{-1}$ was not exceeded in these experiments.[47] Temperature gradient of fluids containing $S_{10Fe}$ and $S_{15Fe}$ increased with increase in concentration of the MGC. For clinical application, the temperature rise of the magnetic fluid should be up to 42 °C. In the presence of a magnetic field, several factors like magnetization, chemical composition, *etc.*, determine the heating capacity of magnetic nanoparticles. The heat generation can be attributed to hysteresis loss and relaxation loss mechanisms. Generally, hysteresis loss occurs in multi-domain ferromagnetic particles with large particle size. In this case, Néel relaxation loss can be ignored, since this relaxation dominates in systems with low coercivity.

Hence, the total heat generation can be attributed to the hysteresis loss and Brownian motion of the nanoparticles. Temperature rise in $S_{10Fe}$ and $S_{15Fe}$ fluids are displayed in Fig. 5 for different MGC concentrations. Initial slope method (ISM) and Box Lucas method (BLM) are two common methods used to calculate SAR value.[48] The SAR value (in W/g) was estimated from the initial linear slope (*dT/dt*) using the ISM relation (4),[48]

$$SAR = C \frac{V_s}{m} \times \frac{dT}{dt} \qquad (4)$$

where *C* (= 4.186 j/g °C) is the specific heat capacity of the solvent, $V_s$ is the sample volume, and *m* is the sample mass.

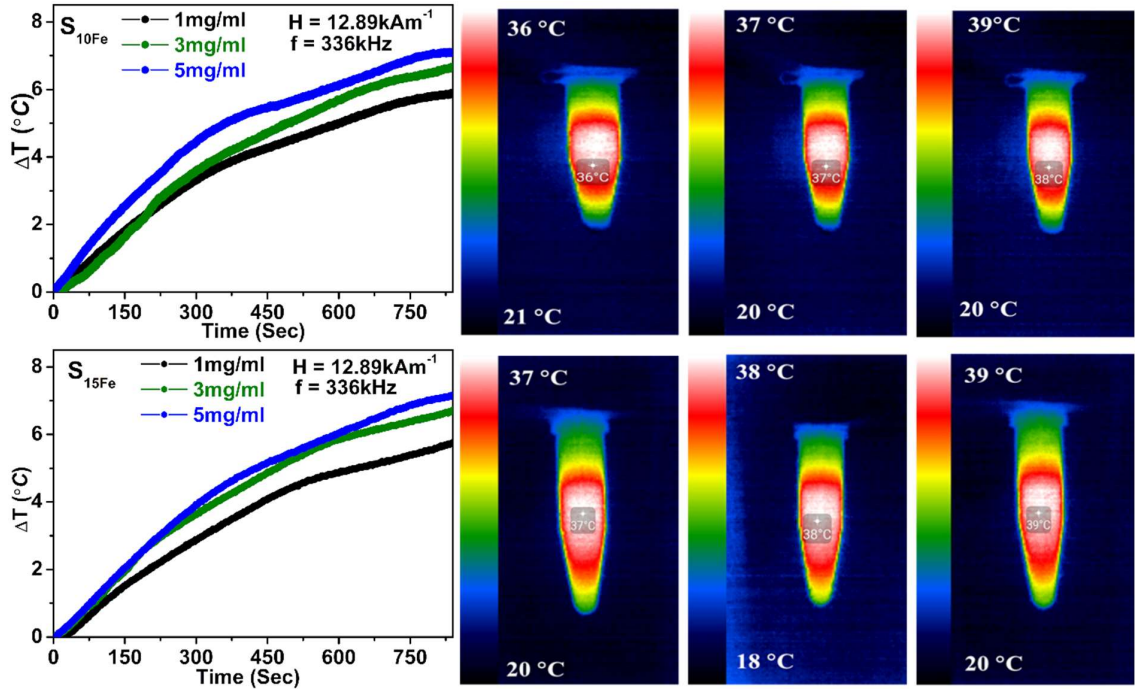

Fig. 5. Curves depicting the temperature raise as a function of time along with near infra-red camera images of induction heated magnetic fluids of $S_{10Fe}$, and $S_{15Fe}$.

BLM was also used to characterize the heating capacity of magnetic fluid by using the relation (5),[49,50]

$$\Delta T = A(1 - e^{-B\Delta t}) \qquad (5)$$

where *A* and *B* are fitting parameters. SAR is defined by relation (6),

$$SAR = \frac{ABC_v m_i}{m_{Np}} \qquad (6)$$

where $C_v$ is the specific heat capacity of solvent, $m_{Np}$ is the mass of magnetic glass-ceramics and $m_i$ is the total mass of the magnetic fluid. SAR value depends on various parameters such as mass of the magnetic fluid, particle size, shape, applied frequency and field strength. Therefore, it is very difficult to compare the SAR value with reported literature. To facilitate comparison, SAR is Therefore, it is very difficult to compare the SAR value with normalized with the applied frequency (*f*), and applied magnetic field amplitude (*H*), to obtain a comparable parameter called the intrinsic loss parameter (ILP), expressed as relation (7).

$$ILP = \frac{SAR}{f \times H^2} \qquad (7)$$

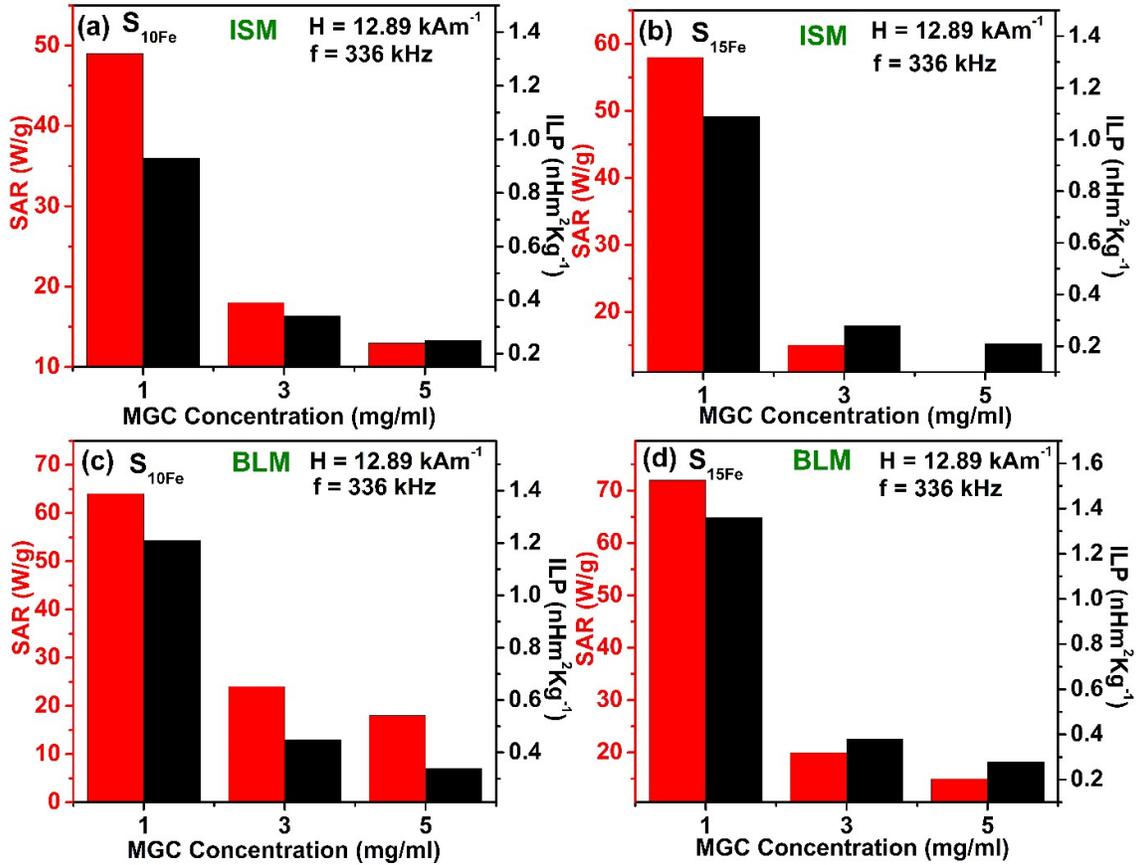

Fig. 6. Calculated SAR and ILP values using ISM for (a) $S_{10Fe}$, (b) $S_{15Fe}$ and BLM for (c) $S_{10Fe}$, and (d) $S_{15Fe}$, for three different concentrations.

Fig. 6 (a) and (b) show the SAR values of $S_{10Fe}$ and $S_{15Fe}$ using ISM while Fig. 6(c) and (d) display the SAR values calculated from BLM for different concentrations of magnetic fluid. It can be seen from the SAR values of $S_{10Fe}$ and $S_{15Fe}$, calculated using ISM are lower than the ones calculated using BLM because ISM considers only the initial linear data and neglects the higher temperature data, whereas BLM considers the data in the whole temperature range.[51] The SAR values show a strong dependence on the MGC concentration in the magnetic suspension in both samples. The magnetic fluid with the lowest MGC content shows the highest SAR value which can be explained through the change in the dipolar interaction between the particles in the fluid.[52].

Table 4. Estimated time for $S_{10Fe}$ and $S_{15Fe}$ fluids to reach 42 °C under $H$ = 12.89 kAm$^{-1}$ and $f$ = 336 kHz.

| Sample ID | Time (sec) | | |
|---|---|---|---|
| | 1 mg/ml | 3 mg/ml | 5 mg/ml |
| $S_{10Fe}$ | 1500 | 1320 | 1280 |
| $S_{15Fe}$ | 1440 | 1300 | 1200 |

When the concentration of magnetic nanoparticles increases, the separation between particles decreases which causes a reduction of heat loss. Ravi *et al.* measured the SAR value of a magnetic fluid with magnetite nanoparticles and graphene oxide as solute.[51] They used a concentration of 0.2 mg/ml which is five times higher than the concentrations of our sample and hence a direct comparison of the two results is not justified. However, the SAR value of $S_{10Fe}$ and $S_{15Fe}$ calculated using BLM for 1mg/ml concentration has an ILP of 1.24 nHm$^2$ kg$^{-1}$ and 1.36 nHm$^2$ kg$^{-1}$, respectively, which is slightly higher than the ILP value of 1.0 nHm$^2$kg$^{-1}$ of commercially available magnetic fluid, FluidMag-CT.[51] When iron oxide was increased from 10 to 15 wt.%, the heating capacity did not increase significantly. This peculiar behavior of heating capacity of $S_{15Fe}$ as compared to $S_{10Fe}$ is a consequence of the stronger magnetic attraction force between magnetic particles in the former, leading to improper heat exchange.

Table 4 displays the estimated time for $S_{10Fe}$ and $S_{15Fe}$ samples to reach the crucial temperature of 42 °C for different magnetic fluid concentrations. It takes 1280 sec and 1200 sec for $S_{10Fe}$ and $S_{15Fe}$ samples at a concentration of 5 mg/ml, respectively, to reach 42 °C from ambient temperature. $S_{15Fe}$ at a concentration of 5 mg/ml took the least time to reach 42 °C. Though $S_{15Fe}$ has better heating capacity over $S_{10Fe}$, has unstable colloidal stability in water. Thus, considering both the ILP and colloidal stability factors, $S_{10Fe}$ has the best potential for HT application.

### 3.6 *In vitro* acellular bioactivity assessment

### 3.6.1. Ion exchange

When the bioactive glass or glass-ceramic comes in contact with the SBF, partial dissolution occurs leading to ionic exchange with the fluid. Generally, bioactive glasses and glass-ceramics degrade with time and the degradation depends upon the concentration and composition of ceramic constituents. Kokubo *et al.* established that the ionic exchange leads to the formation of HAp layer on the ceramic surface, a feature which mimics the interaction of human bone with human blood plasma.[36] The rate of formation of the HAp layer is a measure of the bioactivity of the specimen. Since the HAp layer formation involves ion exchange, pH measurement can be used to probe the mechanism. The change in pH of the SBF containing different MGC for period of 30 days is shown in Fig. 7 (a). Upon immersion of the MGC in SBF, positive ions such as Na$^+$, and Ca$^{2+}$ migrate from the MGC surface *via* exchange with H$_3$O$^+$ ion to the SBF solution. The migration of alkaline metal ions increases the local pH of SBF surrounding the MGC surface.[53] The release of inorganic ions from the specimen surface plays an essential role in stimulating genes involved in tissue regeneration. The bone regeneration ability of bioactive glass or glass-ceramic depends upon the release of Ca and Si species. After 30 days of soaking, the pH of the SBF solution increased from 7.4 to 8.7, 7.4 to 8.5, 7.4 to 8.4, 7.4 to 8.3, 7.4 to 8.2, and 7.4 to 8.0 in the case of $S_{2Fe}$, $S_{4Fe}$, $S_{6Fe}$, $S_{8Fe}$, $S_{10Fe}$, and $S_{15Fe}$ MGC, respectively. Increase in the pH value of MGCs soaked body fluid during the dissolution process is the first response of a bioactive material. The measured pH data of MGC show a sharp increment in pH up to 5 days of soaking in SBF due to the rapid ion-exchange process occurring upon exposure of the MGC pellets to the body fluid. Then the pH values slowly progress toward saturation due to the stabilization of released ions.

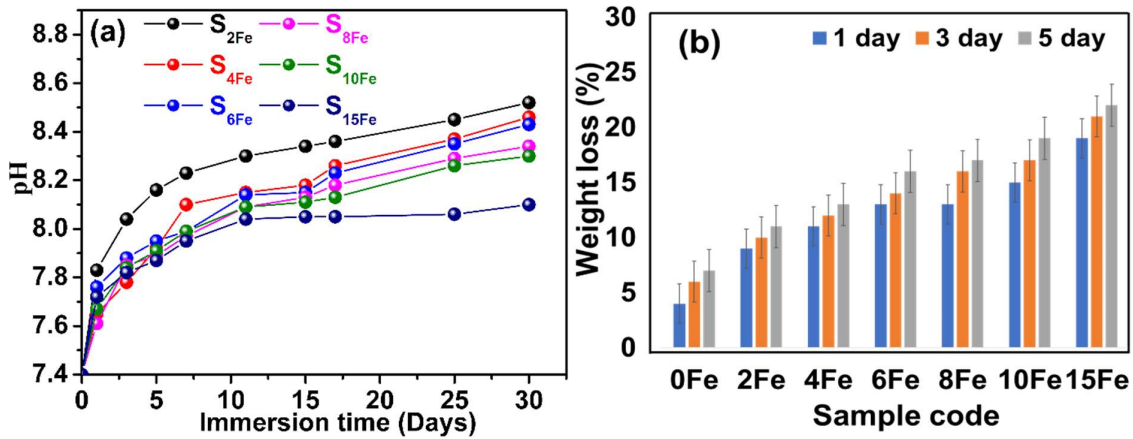

Fig. 7 (a) Change in pH of MGC immersed in SBF as a function of immersion time. (b) Weight loss in MGC measured before and after immersion in PBS for different days.

The observed behavior of time dependent pH data shows that all the MGC exhibit *in vitro* bioactivity expected from an implant material. But, a comparative study of pH curves of different MGC show that the saturation value of pH decreases from 8.7 to 8.0 as iron oxide content is increased. $S_{2Fe}$ shows the maximum variation, which slowly decreases to 8.0 as we move from $S_{2Fe}$ to $S_{15Fe}$. The silanol group (Si-OH) formed in the intermediate stage of the dissolution mechanism is considered to be the nucleating site for a calcium phosphate layer, which can crystallize to form a hydroxy carbonate apatite (HAp) layer. As iron oxide is gradually substituted from the silicon dioxide, the silanol species starts to decrease gradually. This reflects as a decrease in the overall pH of SBF in which the MGC with higher iron oxide content is immersed in. Degradation of the biodegradable material is another competing phenomenon encountered by implants in physiological environment. To understand the hydrolytic degradation of the MGC, MGC powder pellets measuring 10 mm × 10 mm were immersed in the PBS and their weight loss was recorded as a function of time. PBS was chosen as the medium since it is a common medium in cell culture of osteoblast cells. Immersion of bioactive glass-ceramics in PBS follows similar apatite layer formation as observed in SBF before leading to the dissolution process.[54] The PBS treated pellets were dried at 70 °C for 3 hours before measuring their weight. The measured weight loss of the MGC pellets before and after immersion for different days are displayed in Fig. 7 (b).

The weight loss gradually increased with immersion time and a rapid weight loss was observed after 1 day of soaking in PBS due to rapid degradation of the samples. The weight loss also increased with increase in iron oxide content in the Hench's 45S5 composition. These studies indicate that the degradation of glass-ceramics increased with an increase in iron oxide concentration. At the same time, the HAp layer formation also increased which counters the degradation by the deposition of the Ca-P rich layer over the surface of samples. Thus, with increased iron oxide substitution in the Hench's 45S5 composition, slightly decreases in saturation value of pH and a slight increase in degradation is observed. Overall, all MGC show both good ion exchange in SBF and degradation in PBS which qualify them as degradable bioactive materials.

### 3.6.2. Surface apatite layer formation

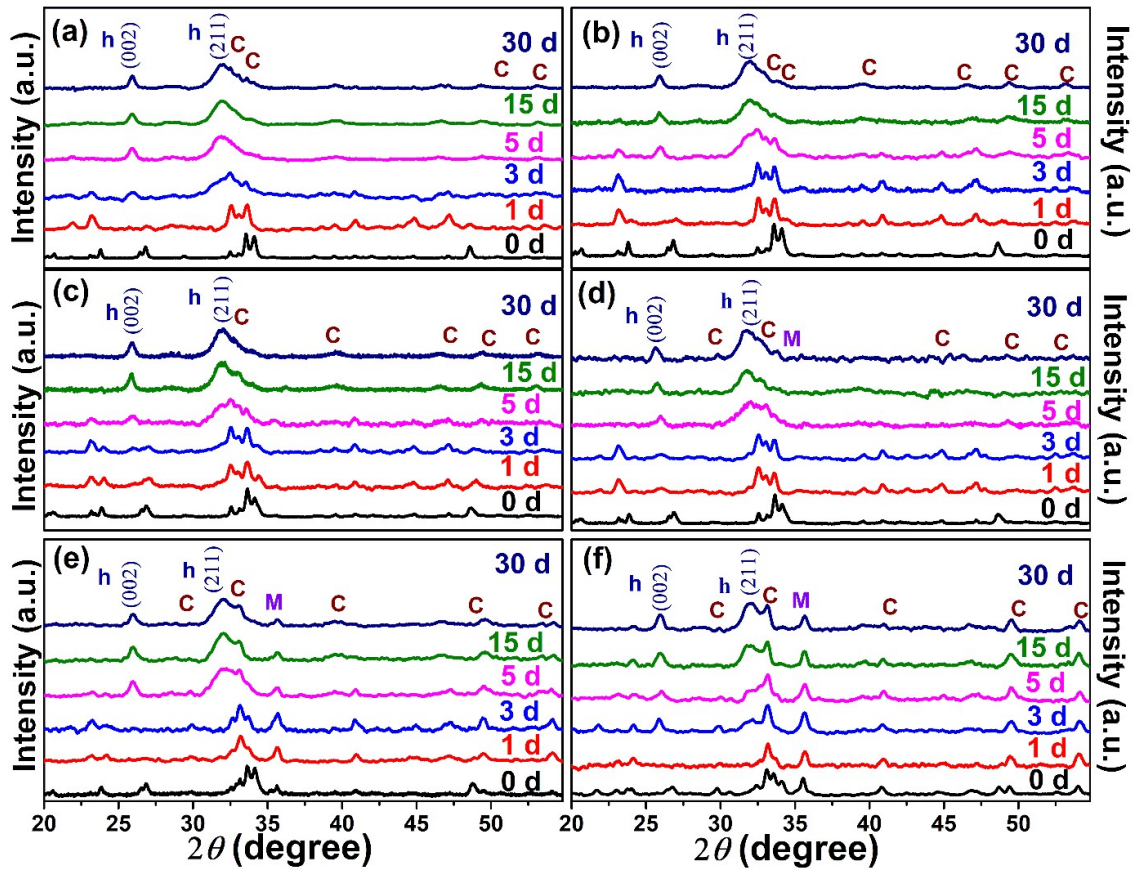

Fig. 8. GI-XRD patterns of (a) S$_{2Fe}$, (b) S$_{4Fe}$, (c) S$_{6Fe}$, (d) S$_{8Fe}$, (e) S$_{10Fe}$, and (f) S$_{15Fe}$ pellets immersed in SBF for different days. Symbols M, h and C represent magnetite, hydroxyapatite, and combeite phases, respectively.

*In vitro* acellular bioactivity test was carried out by soaking MGC pellets in SBF maintained at initial pH of 7.4 and temperature of 37 °C. The structural change occurring on the surface of MGCs after soaking in SBF was observed by recording grazing incidence XRD (GI-XRD) patterns at regular time intervals. Fig. 8 (a) to (f) show XRD patterns of SBF treated MGC pellet surfaces recorded after 0, 1, 3, 5, 15, and 30 days of immersion. The XRD pattern labelled as 0 days corresponds to the MGC before immersion in SBF, which exhibits the crystalline nature of the samples with magnetite, hematite, and combeite phase, as already discussed in the section on structural analyses of the MGCs. After immersion in SBF, a new crystalline phase is observed in the XRD pattern indicating a crystalline layer formation on the surface of the MGC. Two well-defined XRD peaks at 2$\theta$ values ~26° and ~32° develop on the surface of S$_{2Fe}$ pellet after three days of soaking in SBF.

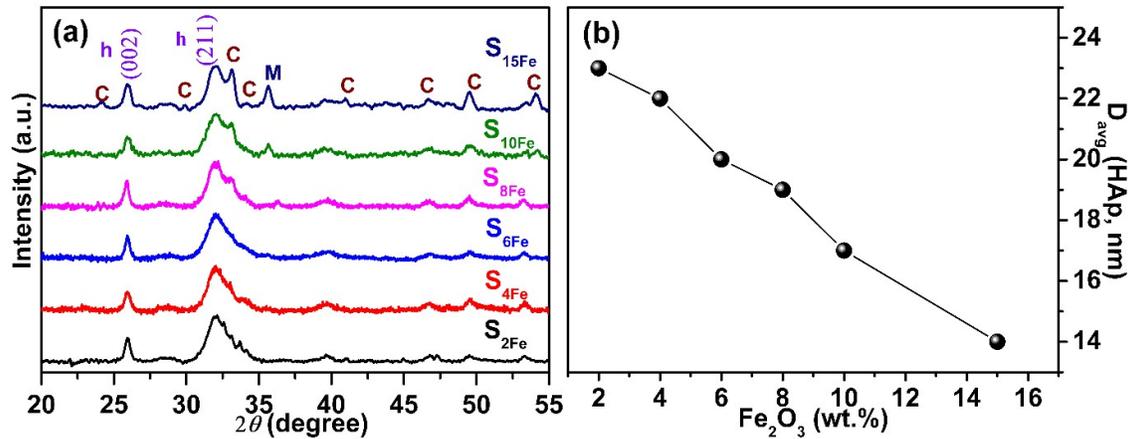

Fig. 9. (a) XRD patterns, and (b) average crystallite of HAp layer in MGC after 30 days of soaking in SBF.

These XRD peaks appear only after five days of soaking in SBF in $S_{4Fe}$, $S_{6Fe}$, $S_{8Fe}$, $S_{10Fe}$, and $S_{15Fe}$. These two prominent peaks could be assigned to reflections from (002) and (112) planes of HAp crystallites (ICDD 74-0565). Formation of the bone mineral phase HAp formation on the surface of the samples has already been explained through the ion exchange mechanism upon soaking in SBF (section 3.6.1). The intensities of the newly developed crystalline (HAp) phase peaks increase with immersion time as displayed in Fig. 8 (a) to (f). This increase in HAp phase with immersion time can be elucidated by the accumulation of $Ca^{2+}$ and $PO_4^-$ ions on the surface of MGCs. This HAp layer formation and increase in pH which was discussed in section 3.6.1 combinedly indicate the bioactive nature of all MGC compositions. Fig. 8 (a-f) indicate a gradual development of the HAp layer with immersion time (days) and gradual suppression of the intensity of the underlying crystalline phases of each MGC. However, the crystallization of HAp layer depends upon the concentration of Ca and P ions and the pH of the solution. However, the thickness and the surface coverage of the deposited HAp layer decrease in MGC as their iron oxide is increased. One can see that the HAp layer thickness and coverage over $S_{8Fe}$, $S_{10Fe}$ and $S_{15Fe}$, are low and thus unable to completely suppress the appearance of the underlying glass-ceramic peaks in the GI-XRD patterns even after 30 days of immersion. Fig. 9 (a) exemplifies the fact that the HAp layer is progressively thinner on $S_{8Fe}$, $S_{10Fe}$ and $S_{15Fe}$ surfaces indicating a decrement in the bioactivity in MGC with higher iron oxide content. Fig. 9 (b) shows the average crystallite size of HAp decreases almost linearly with increase in iron oxide content. Hence, one can infer that the HAp layer growth (a) increases with immersion time for each MGC and (b) slightly decreases for MGC with higher iron oxide concentration. The relative decrease of bioactivity of iron substituted glass-ceramic with increase in iron oxide content could be explained in terms of variation in the crystalline phase percentages estimated in the XRD analysis of the MGCs (Fig. 2 (a)). Substitution of iron oxide into the Hench composition increases the number of $Fe^{2+}$ and $Fe^{3+}$ ions at the cost of $Si^{4+}$ ions. A reduction in $Si^{4+}$ ions decreases the silanol group formation in SBF which in turn decreases the HAp layer formation capacity or bioactivity of the MGC.

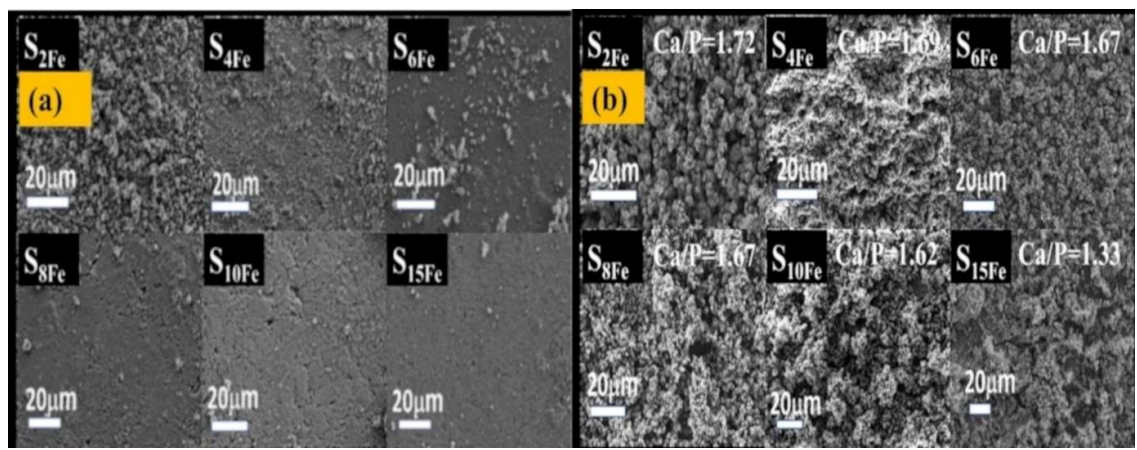

Fig. 10. FESEM images of MGC after immersion in SBF for (a) 5 days and (b) 30 days.

Fig. 10 (a) and (b) show the FESEM images of the surface of the MGC after 5 and 30 days of immersion in SBF which give a visual evidence for the surface layer growth. The grainy white layer signifies the HAp layer formed on the surface of immersed glass-ceramics. After soaking for five days, high density of spherical HAp grain formation is observed on the surface of $S_{2Fe}$ and $S_{4Fe}$ samples. With increasing immersion time, HAp layer growth increases, and after 30 days of immersion, the layer covers the whole surface of the specimen (Fig. 10 (b)). Rapid layer formation of $S_{2Fe}$ and $S_{4Fe}$ describes the faster interchange of $Ca^{2+}$ ions of a specimen to the $H_3O^+$ ions from a physiological fluid. Due to the interchange of ions, silanol group Si-OH is formed on the surface of the pellets, which induces HAp nucleation. In addition, the rate of HAp layer formation mainly depends upon the composition of the MGC. The surface mineralization was confirmed by EDS analysis, which shows the Ca/P ratio for all the samples after 30 days of soaking (*c.f.* Fig. 10 (b)). The Ca/P ratio for $S_{4Fe}$, $S_{6Fe}$, $S_{8Fe}$, and $S_{10Fe}$, are close to the Ca/P ratio of 1.67 of reported for bulk HAp. The lower Ca/P ratio of the $S_{15Fe}$ sample suggests lower bioactivity as also supported by the pH and XRD analyses of the sample.

### 3.6.3. Osteoblast cell viability

Cell viability tests were performed to understand the effects of iron oxide substitution on the growth and proliferation of osteoblast cells on the Hench's 45S5 glass-ceramic surface. Cytocompatibility studies were also carried out to determine the optimally iron oxide substituted MGC with appropriate magnetic properties and bioactivity required for biomedical applications. The observed results of osteoblast (MG-63) cell viability tests are shown in Fig. 11 (a) to (c). For the tested concentration range of 2 mg/ml to 0.25 mg/ml of Hench's 45S5 glass-ceramic without iron oxide, i.e., $S_{0Fe}$, the osteoblast cells showed excellent cytocompatibility and enhanced cell proliferation for 01 and 03 days of incubation as expected. MG-63 cells showed viability varying from 91% to 101% for the $S_{0Fe}$ concentration range of 2 mg/ml to 0.25 mg/ml incubated for 01 day. For 03 days of incubation, the cell viability increased to 92% to 111% for the same concentration range of $S_{0Fe}$ sample. The increment in cell viability can be attributed to the components of MGC such as Ca, P and Si which reportedly promote bone cells growth and proliferation.[55] Ca and P are key components of natural bone and Ca/P ratio of the MGC mimics the natural scaffold for osteoblast growth and proliferation.

Ca supports the mineralization of extracellular matrix of bone and also increases the expression of growth factors such as IGF-I and IGF-II.[56]

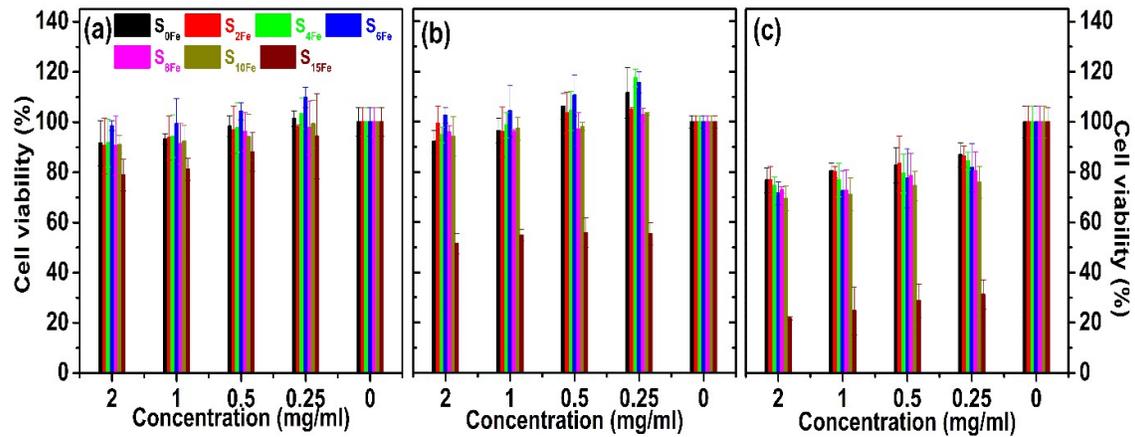

Fig. 11. Viability of MG-63 human osteoblast cells incubated with MGC for (a) 01 day, (b) 03 days, and (c) 05 days.

Si plays an essential role in cellular metabolic processes like bone tissue formation and calcification. Intake of Si also increases the bone mineral density and stimulates the collagen I formation.[57,58] These factors are responsible for the increased viability observed in case of MG-63 cells incubated with $S_{0Fe}$. Further, 5 days of incubation with $S_{0Fe}$ showed decreased viability of 77% (2 mg/ml) to 87% (0.25 mg/ml). This decrease in cell viability can be attributed to the reduction in bioavailability of $S_{0Fe}$ due to its biodegradation with time and over confluence of the culture plate well, leading to death of the osteoblast cells.[59] The multifunctional MGC incorporated with 2, 4, 6, 8, 10 and 15 wt.% iron oxide by replacing silicon oxide were also tested for their cytocompatibility and osteoblast cell growth for the same sample concentrations as done with $S_{0Fe}$. After 01 day of incubation, osteoblast cells showed none or slight increase in viability for the concentration range of 2 to 0.25 mg/ml of various MGC. After incubation of 03 days, the cell viability of osteoblast was observed to significantly increase in the presence of these MGCs. For $S_{2Fe}$ sample concentration of 2 to 0.25 mg/ml cell viability increased by 99% to 105%. Similar increment in the cell viability was observed for the $S_{4Fe}$ and $S_{6Fe}$ samples. However, $S_{8Fe}$ and $S_{10Fe}$ showed a slight decrease in the cell growth between 94 to 103% for the tested concentrations. In contrast, $S_{15Fe}$ containing 15 wt.% iron oxide showed only 55% cell viability indicating the toxic nature of the sample. The enhanced cell viability for 2 to 10 wt.% of iron oxide MGC samples was attributed to the presence of Fe in the medium. Along with the desired magnetic behavior, Fe reportedly stimulates the growth and proliferation of osteoblast cells and thus increased cell viability was observed in the MGC with iron oxide phases.[60,61] Further, as discussed above, the incorporation of Fe results in proportionate decrease in Si, which lowers the collagen formation and thereby limiting the cell proliferation rate. In sample $S_{15Fe}$, the reduced amount of Si may have resulted in the decreased bioavailability of MGC, thus leading to significant reduction in cell viability. After 05 days of incubation, the cell viability for all the samples, barring $S_{15Fe}$, decreased up to 70% for the highest concentration of 2 mg/ml of MGC. Such significant decrease in the cell number signifies extensive biodegradation of the MGC resulting in significant reduction in bioavailability of bioactive components like Ca, P and Si. Also, over confluence of culture wells might have affected the cell viability. In the case of $S_{15Fe}$, the already low cell viability observed after 03 days of incubation period was further lowered to 30%.

**Conclusions**

Mesoporous MGC with compositions of (45-x) $SiO_2$ 24.5CaO 24.5$Na_2O$ 6$P_2O_5$ x$Fe_2O_3$ (0 ≤ x ≤ 15) have been successfully synthesized by sol-gel route by substituting iron oxide for silicon oxide in the Hench's 45S5 bioactive glass composition. The surface area of the porous MGCs increases with iron oxide content, while the pore size and pore volume increase for samples with 8 wt.% iron oxide, and then show a decrease. XRD patterns of sol-gel derived MGC powder reveal the presence of three crystalline phases in all the investigated samples. The combeite phase percentage decreased and the magnetic (hematite and magnetite) phases increased with increased iron oxide substitution. The magnetization of the MGC increased with increase in iron oxide substitution. Induction heating studies reveal the maximum SAR value (calculated by BLM) for the MGC sample with the highest iron oxide content which also was the quickest (within 1280 s) to reach 42 °C from room temperature under test conditions. The *in vitro* acellular bioactivity study of the MGC shows a rapid degradation of formed HAp layer due to decrease in silanol group with an increase in iron oxide substitution. All samples except $S_{15Fe}$ were found to be non-toxic to MG-63 osteoblast cells. This study shows a decreasing trend in bioactivity and increasing trend in induction heating capacity with an increase in iron oxide content in the MGCs. MGC sample $S_{10Fe}$ exhibiting MG-63 cell viability of 87% after 5 days of incubation, $M_s$ of 0.82 emu/g, $H_c$ of 149 Oe, ILP of 1.24 $nHm^2\ kg^{-1}$ has all the features to be a promising thermoseed for hyperthermia treatment of cancer when compared to the ILP and bioactivity of commercially available thermoseed fluid, FluidMag-CT. It is to be noted that these nanoporous MGC powders have the same CaO, $Na_2O$, and $P_2O_5$ content as Hench's 45S5 bioglass-ceramics, thereby providing them with good induction heating capability and mechanical strength without compromising on bioactivity and biodegradability required for bone regeneration.

**Author Contributions**

Nitu prepared all samples, executed the measurements, analysed the data, and prepared the original manuscript draft. R. R Fopase and L. M. Pandey help to investigate the cell assessment of prepared samples. P. Seal and J. P. Borah help to collect induction heating data. A. Srinivasan conceptualized the idea and supervised the above study. Also, A. Srinivasan edited and approved the final manuscript.

**Conflicts of interest**

There are no conflicts to declare.

**Acknowledgements**

The authors thankfully acknowledge utilization of facilities in Department of Physics, IIT Guwahati acquired under the Department of Science and Technology, Government of India project no: SR/FST/PSII-037/2016, Central Instrument Facility, Centre for the Environment, IIT Guwahati. Nitu thanks, the Ministry of Education, India for a fellowship to pursue Ph.D. The author also thanks to Rajendra K Singh for collecting NIR images.